\documentclass{aa}
\usepackage{graphicx}
\begin{document}
   \title{An improvement of the BeppoSAX LECS and MECS \\ positioning accuracy}

   \author{M. Perri
          \inst{1,2}
          \and
           M. Capalbi\inst{1}
          }

   \offprints{M. Perri, \email{perri@asdc.asi.it}}

   \institute{ASI Science Data Center, c/o ESA-ESRIN, 
	      via Galileo Galilei, I-00044 Frascati, Italy\\
              \email{perri@asdc.asi.it, capalbi@asdc.asi.it}
         \and
	       Dipartimento di Fisica, Universit\`a ``La Sapienza'', 
               Piazzale A. Moro 2, I-00185 Roma, Italy\\
             }

   \date{Received / Accepted }

   \abstract{
We present a study of the source positioning accuracy of the LECS and MECS instruments on-board {\it Beppo}SAX.
From the analysis of a sample of archival images we find that a systematic error, which depends on
the spacecraft roll angle and has an amplitude of $\sim$ 17'' for the LECS and $\sim$ 27'' 
for the MECS, affects the sky coordinates derived from both instruments.

The error is due to a residual misalignment of the two instruments with respect to the 
spacecraft Z axis arisen from the presence of attitude inaccuracies in the observations used to 
calibrate the pointing direction of LECS and MECS optical axes.

Analytical formulae to correct LECS and MECS sky coordinates are derived. After the coordinate 
correction the 90\% confidence level error radii are 16'' and 17'' for LECS and MECS respectively, improving 
by a factor of $\sim 2$ the source location accuracy of the two instruments.
The positioning accuracy improvement presented here can significantly enhance the follow-up studies at other 
wavelengths of the X-ray sources observed with LECS and MECS instruments.

   }

   \authorrunning{M. Perri \& M. Capalbi}

   \maketitle


\section{Introduction}
\label{intro}

The 0.1-10 keV imaging instruments on-board the {\it Beppo}SAX satellite (Boella et al. \cite{boella97a}) 
contribute significantly to the study of the X-ray sky.

Among the important topics studied, we can certainly include the investigation of the nature of 
the sources producing the Cosmic X-ray Background (Giommi et al. \cite{giommi98}, \cite{giommi00}; 
Fiore et al. \cite{fiore99}, \cite{fiore01}; Comastri et al. \cite{comastri}; Vignali et al. \cite{vignali}; 
La Franca et al. \cite{lafranca}), the study of the X-ray afterglows of Gamma-Ray Bursts 
(e.g., Antonelli et al. \cite{lucio}; Feroci et al. \cite{feroci}; in 't Zand et al. \cite{zand01}; 
Piro et al. \cite{piro}) and the identification of new X-ray transients in the galactic bulge (e.g., 
the galactic black hole candidate XTE J1908+094, in 't Zand et al. \cite{zand02}).

Crucial, in all these studies, is an accurate determination of the celestial coordinates of the imaged 
X-ray source, since it may allow a firm identification of its counterpart at other wavelengths.

Previous discussions of the positioning accuracy of the {\it Beppo}SAX MECS instrument 
(Ricci et al. \cite{ricci}; Fiore et al. \cite{fiore01}) reported an error of $\sim$ 1 arcmin at the 
90\% confidence level and pointed out that a significant contribution to the X-ray coordinate error comes 
from systematic uncertainties due to the absolute spacecraft attitude reconstruction.

With the aim of understanding the origin of these systematic uncertainties we present here a detailed 
study of the source location accuracy of the {\it Beppo}SAX imaging instruments.


\section{Data analysis}
\label{data}

The scientific instrumentation on-board the {\it Beppo}SAX satellite includes four co-aligned 
X-ray telescopes, each composed of a grazing incidence Mirror Unit and of a position sensitive 
Gas Scintillation Proportional Counter located at the focal plane.

Three of these systems are nearly identical and, collectively, constitute the Medium Energy Concentrator Spectrometer 
(MECS, Boella et al. \cite{boella97b}). The three units, named MECS1, MECS2 and MECS3, are sensitive in the 
1.3-10 keV energy band.

\begin{figure*}
  \centering
  \includegraphics[width=8.2cm]{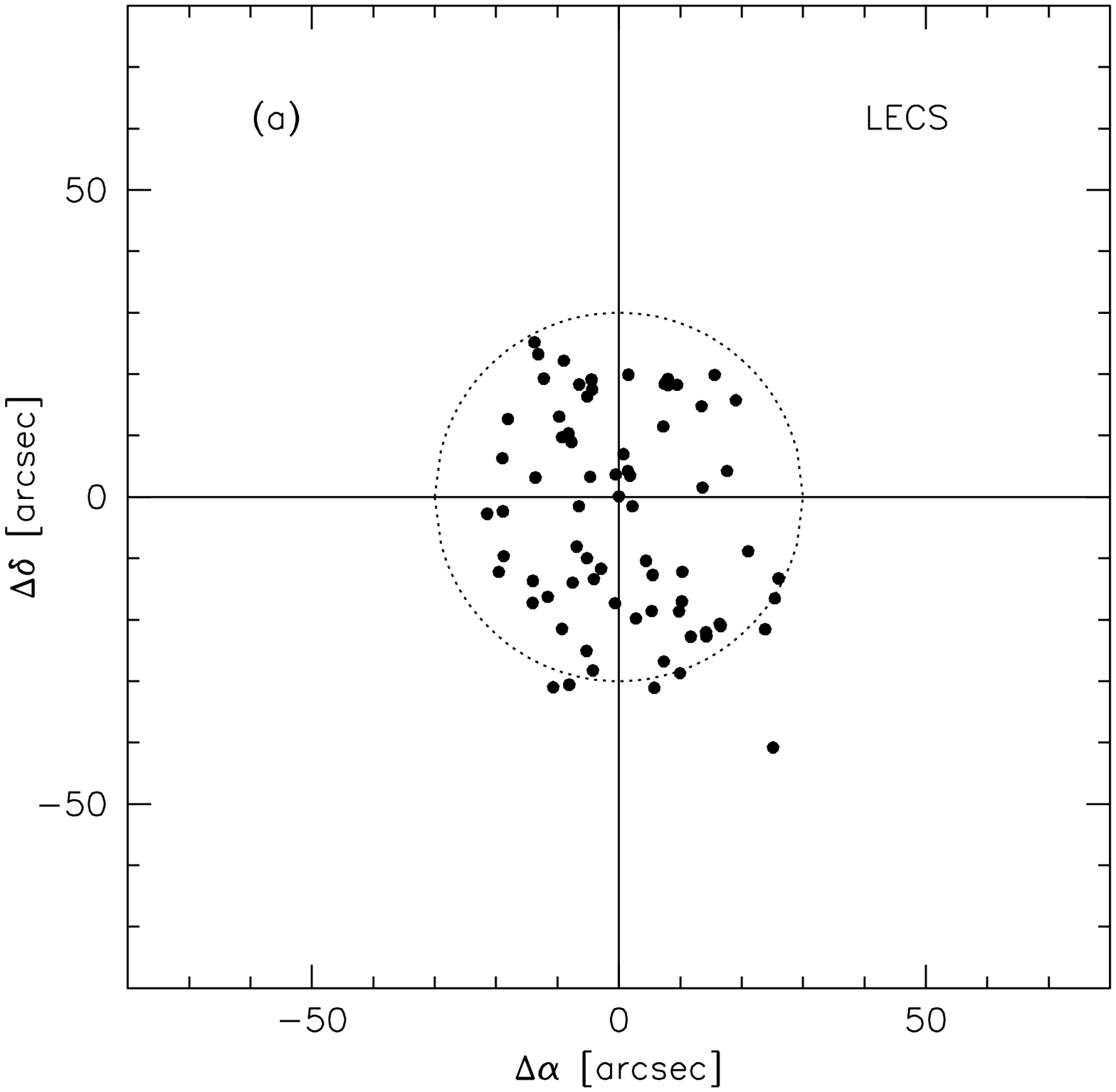}
  \includegraphics[width=8.2cm]{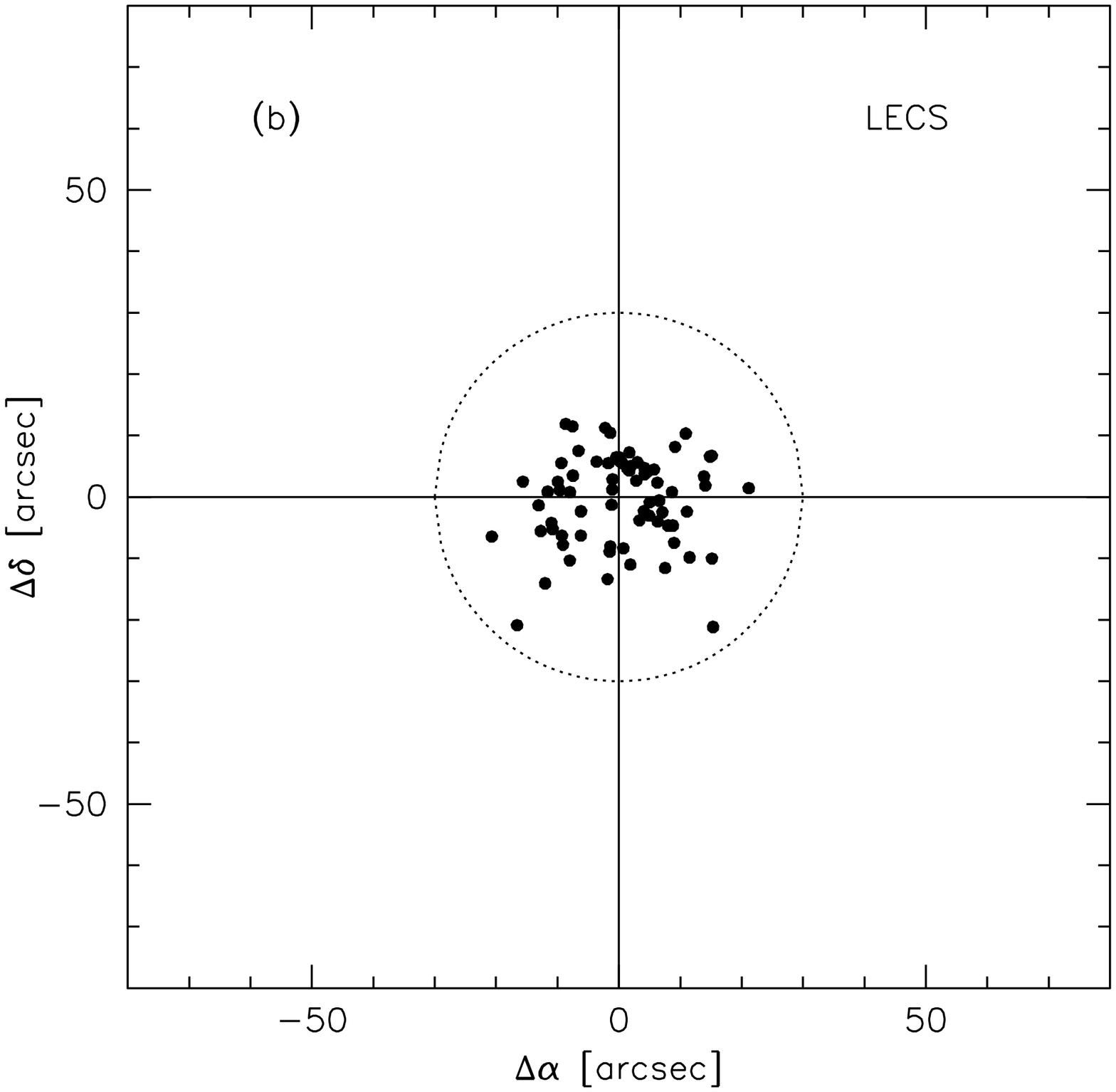}
  \includegraphics[width=8.2cm]{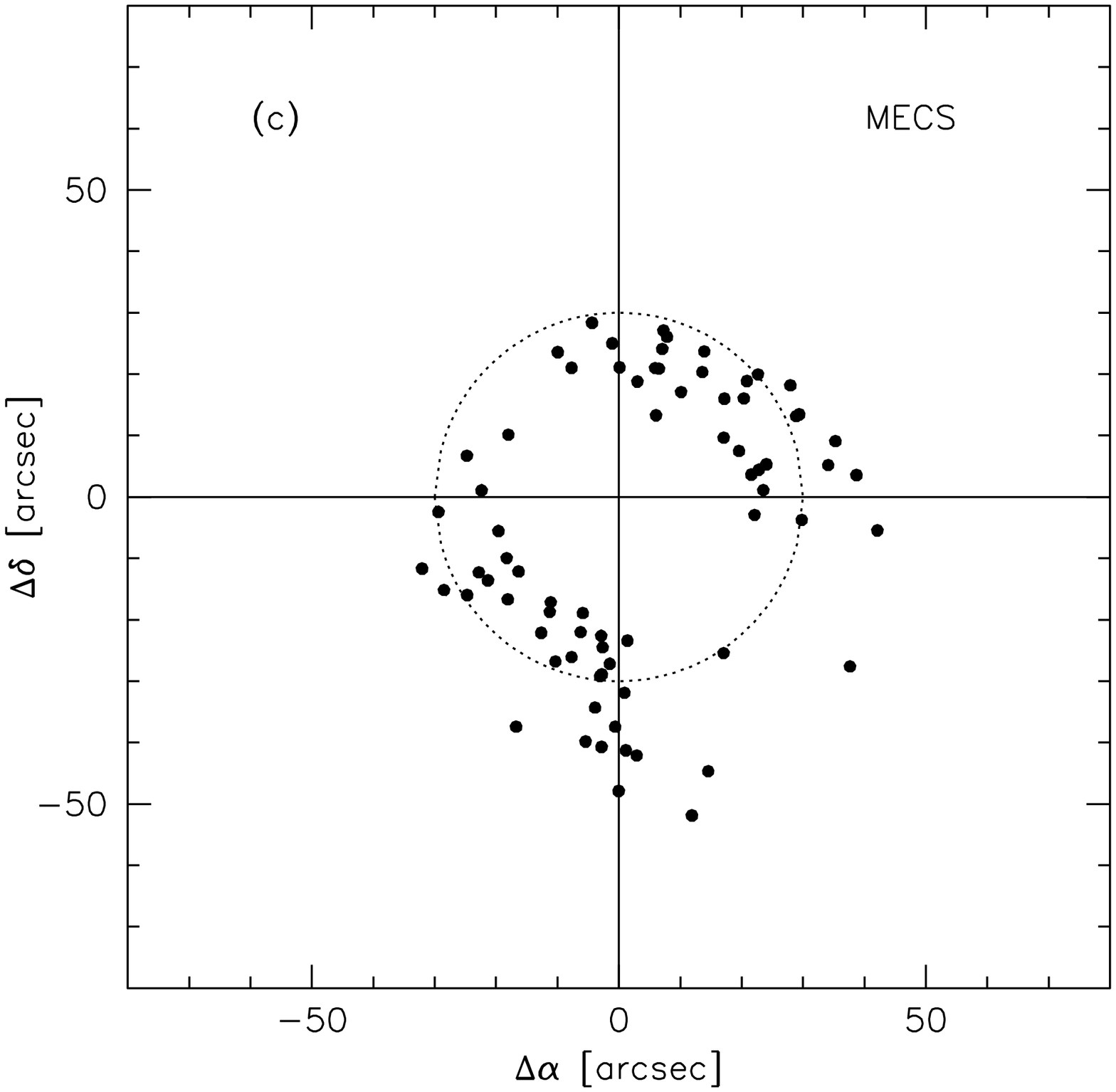}
  \includegraphics[width=8.2cm]{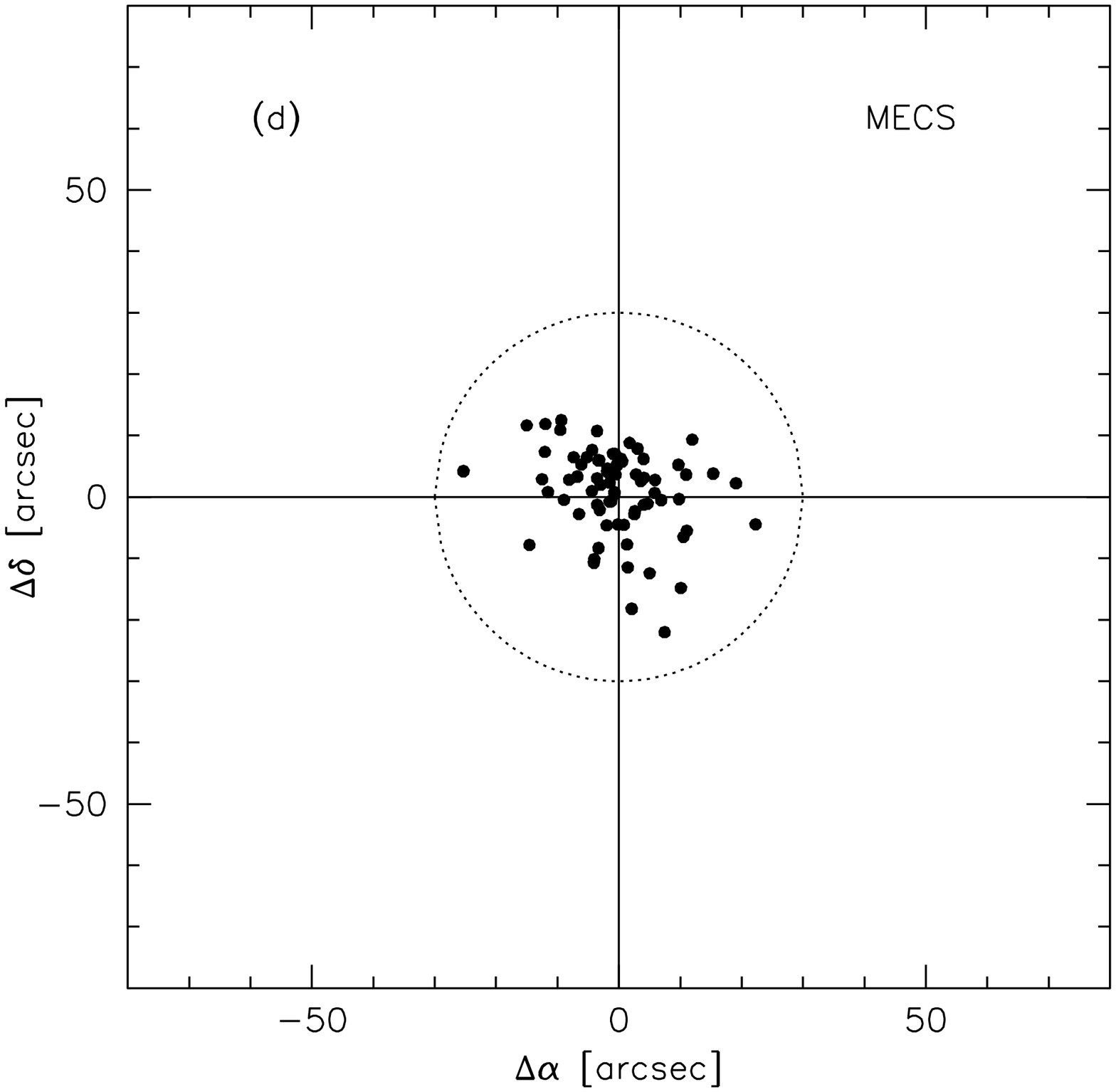}
  \caption{ 
	    {\bf a)} The deviation in RA and Dec between LECS and optical 
		     positions for observations performed after May 6th 1998. 
	    {\bf b)} The same plot after the coordinate correction discussed in 
		     the text.
	    {\bf c)} The deviation in RA and Dec between MECS and optical 
		     positions for observations performed after May 6th 1998.
	    {\bf d)} The same plot after the coordinate correction discussed in 
		     the text. The dotted circles have a radius of 30''.
  }
  \label{circles}
\end{figure*}

The fourth system, the Low Energy Concentrator Spectrometer (LECS, Parmar et al. \cite{parmar}), 
has a mirror design identical to the one of the other three units, but the detector is sensitive 
to X-ray photons in the 0.1-10 keV energy band.

The image size of both instruments is $256 \times 256$ pixels and the detector pixel size, near the 
center of the field of view, is $\sim 14$'' for the LECS and $\sim 19$'' for the MECS.

We have studied the source positioning accuracy of LECS and MECS instruments using a set of X-ray sources 
observed by {\it Beppo}SAX for which the position of the optical counterpart is known to within 1''.

LECS and MECS fields have been selected according to the following criteria:
\begin{itemize}
\item the target is on-axis and relatively bright \\
($> 10^{-2} ~ \mathrm{cts~s}^{-1}$ for both LECS and MECS)
\item the target is neither extended nor confused
\item the whole range of spacecraft roll angle values, 
from $-90^\circ$ to $+270^\circ$, is uniformly covered (for a definition of roll angle see 
the Appendix)
\item the observation has been performed later than May 6th 1998.
\end{itemize}

The last condition has been imposed to minimize position uncertainties since the {\it Beppo}SAX 
attitude data before May 6th 1998 are affected by relatively large inaccuracies 
(see Sect.~\ref{origin}).

The above conditions resulted in the selection of a sample of 72 X-ray fields which consists of 
49 pointings of AGNs, 14 of Stars and 9 of X-ray binaries\footnote{A table containing the 
list of the sources used in this analysis is available at 
http://www.asdc.asi.it/bepposax/positions/table.html.}.

LECS and co-added MECS2+MECS3 (hereafter MECS) images have been used. MECS1 data are not included 
in the analysis since this unit failed on May 6th 1997.

Images  were accumulated in the 2-10 keV band (channels 44-220) 
for the MECS and in the 0.1-9.5 keV band (channels 10-950) for the LECS. Event files 
from the {\it Beppo}SAX public archive (Giommi \& Fiore \cite{pao_fab}) at the 
ASI Science Data Center (ASDC) have been used.

To determine the celestial coordinates of the X-ray sources we used the sky positions 
stored in the headers of LECS and MECS event files.
The source detection has been performed using a variation of the DETECT routine of the XIMAGE package 
(Giommi et al. \cite{giommi92}) as described in Fiore et al. \cite{fiore01}.

Special attention has been devoted to the quality of the detections and each source position 
has been visually checked.


\section{Comparison with optical coordinates}
\label{comp}

We have compared the celestial coordinates of the X-ray sources derived from LECS and 
MECS images with the accurate positions obtained from optical catalogs 
(Veron-Cetty \& Veron \cite{veron}; Turon et al. \cite{turon}; van Paradijs 
\cite{xrbcat}).

We have then computed RA and Dec offsets between the optical and LECS/MECS values (labelled respectively 
``L'' and ``M'') using the following definitions:
\begin{eqnarray}
	\label{d_ra}
	\Delta \alpha_\mathrm{L,M} = (\alpha_\mathrm{opt}-\alpha_\mathrm{L,M})~ \mathrm{cos}\,\delta ~~\\
	\label{d_dec}
	\Delta \delta_\mathrm{L,M} = \delta_\mathrm{opt} - \delta_\mathrm{L,M} 
              ~~~~~~~~~~~~~
\end{eqnarray}

Note that RA offsets have been corrected by the factor $\mathrm{cos}\,\delta$ and represent 
therefore the true separation in the sky.

In Fig.~\ref{circles} (panels a and c) we plot the measured $\Delta \alpha$ vs $\Delta \delta$ values for 
the LECS and MECS instruments.

We have next computed the angular distances between the X-ray and optical positions. 
In Table~\ref{tab} (top) we list the ``radius'' within which 68\% and 90\% of the objects 
are included. All the values in Table~\ref{tab} have been rounded to unity.


\begin{table} [h]
   \caption{LECS and MECS (2 units) 68\% and 90\% error radius values before and after the coordinate 
correction of Eqs.~(\ref{ra_cor}) and (\ref{dec_cor}) for observations performed after May 6th 1998.}
   \label{tab}
     \begin{tabular}{cccc}
     \hline
     \hline
  &   ~Instrument~ & ~68\% radius~ & ~90\% radius~ \\
     \hline
	&	 &                   &                   \\
before   &  LECS & 23'' & 29'' \\
~correction~   &  MECS & 30'' & 41'' \\
	&	 &                   &                   \\
     \hline
	&	 &                   &                   \\
after   &  LECS & 12'' & 16'' \\
correction   &  MECS & 11'' & 17'' \\
	&	 &                   &                   \\
     \hline
     \end{tabular}
\end{table}

\section{The dependence on roll angle}
\label{roll_ang}

As can be seen in Fig.~\ref{circles} (panel c) the $\Delta \alpha$ and $\Delta \delta$ values for 
the MECS are not uniformly distributed around the zero values, indicating that 
a systematic error affects the source positioning accuracy of the instrument. 
For the LECS (panel a) we note that the same effect is not evident.

Moreover, Table~\ref{tab} shows that the positions derived from the LECS 
instrument are more accurate (29'' error radius at the 90\% confidence level) with respect 
to those obtained from the MECS (41'').

To investigate in more details these results we have searched for a possible dependence of RA and 
Dec offsets on the satellite roll angle $\rho$.

\begin{figure*}
  \includegraphics[width=18.4cm]{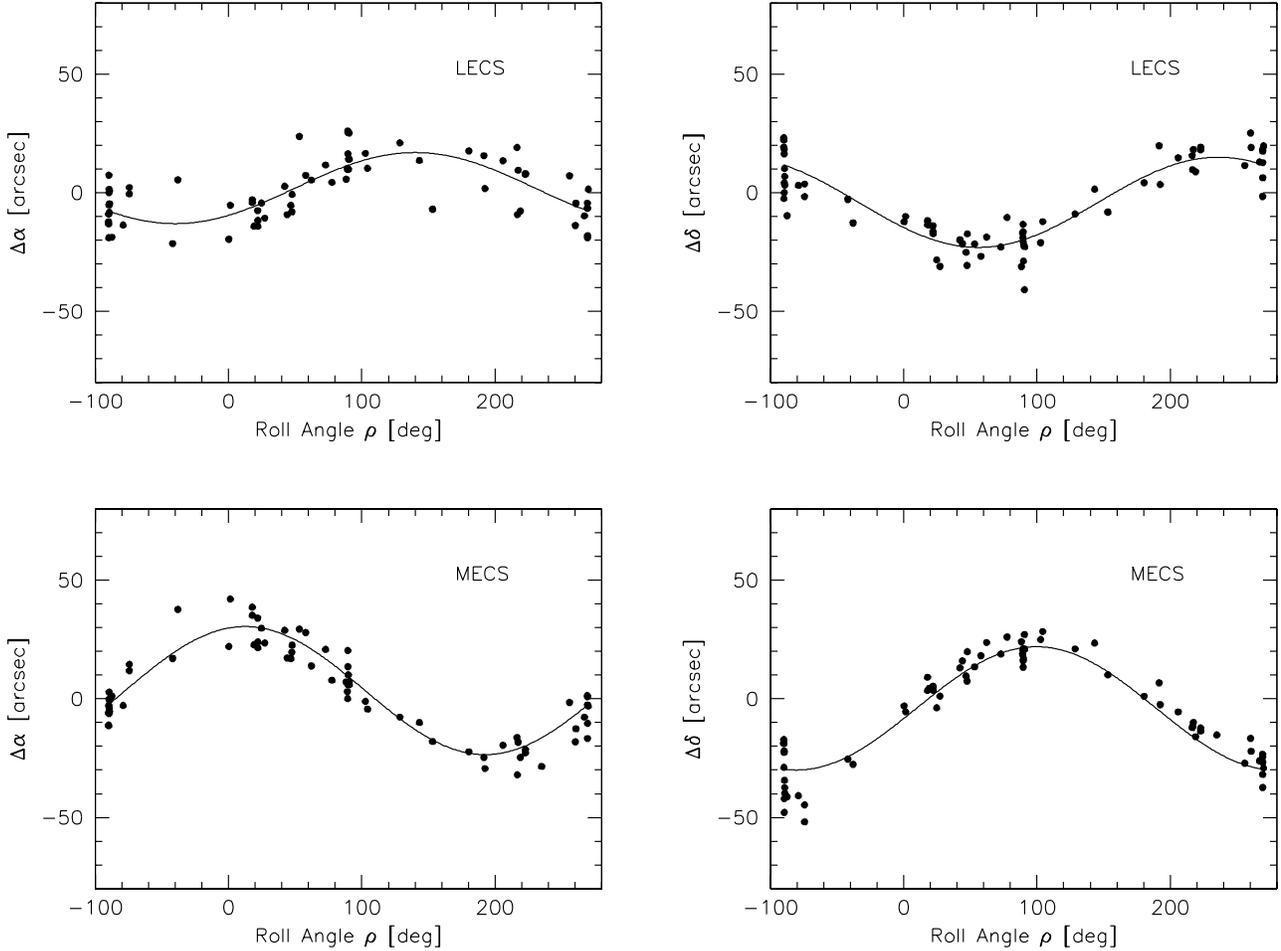}
  \vspace*{-5cm}
  \caption{Top: The differences in RA and Dec between LECS and optical 
	        positions plotted as a function of the spacecraft roll angle. 
	   Bottom: Same plot for the MECS instrument.
           The solid lines are best fits to the data and are discussed in the text.}
  \label{roll}
\end{figure*}

In Fig.~\ref{roll} we plot $\Delta \alpha$ and $\Delta \delta$ as a function of the spacecraft roll angle. 
As can be clearly seen, for both instruments a strong correlation between these quantities and the roll 
angle is found. Furthermore, we see that $\Delta \alpha$ and $\Delta \delta$ values follow a sinusoidal law 
and that the amplitude of this effect is $\sim$ 15'' in the case of the LECS and $\sim$ 25'' for the MECS.

We have also studied individually the single MECS units to verify if a specific offset 
dependence on the roll angle is present. We performed the analysis described 
above on MECS2 and MECS3 images separately obtaining the same results found 
in the study of the co-added images\footnote{An analysis of the positioning accuracy of the 
single MECS units can be found at http://www.asdc.asi.it/bepposax/report/report.html.}.

The fact that LECS and MECS coordinate offsets are correlated with the satellite 
roll angle is a sure indication of a residual misalignment of both instruments with respect to 
the spacecraft Z axis (the one co-aligned with the instruments).

The origin of the residual misalignment of LECS and MECS instruments is discussed in the next section.


\section{The LECS and MECS residual misalignment}
\label{origin}

As already mentioned in Sect.~\ref{data}, the {\it Beppo}SAX attitude data 
during the first two years of the mission (June 1996-May 1998) have been 
affected by relatively large systematic inaccuracies. 

Since, as we will see, these inaccuracies are at the origin of the observed LECS and MECS 
residual misalignment, we briefly discuss this topic. Some useful 
definitions and formulae concerning the attitude can be found in the Appendix.

In summer 1997 an error of about 20'' in the misalignment matrices of the three 
satellite star-trackers, used to track the guide stars and control the 
spacecraft attitude, was discovered. 

As the following investigation revealed, the error originated because the coordinates 
of the guide stars used by the spacecraft star-trackers were not corrected for the
annual aberration due to the heliocentric motion of the Earth.

A new computation of the star-trackers misalignment matrices was soon performed (August 1997), 
but during the following months rather large movements of the 
spacecraft (up to $\sim 2$ arcmin) when the attitude control was switched between 
different star-trackers were observed.

A detailed analysis of the problem pointed out an error in the correction for annual 
aberration of the guide stars coordinates which caused even larger errors in the 
star-trackers misalignment matrices and, consequently, larger attitude inaccuracies. 
A new and exact computation of the star-trackers misalignment matrices was then 
performed on May 6th 1998.

The problems listed above have determined spacecraft attitude data inaccuracies for 
observations carried out in the period June 1996 - May 6th 1998. Due to both technical and 
financial reasons, a new computation of the attitude of this set of observations is not foreseen.

We identify these attitude inaccuracies as the cause of the observed residual 
misalignment of LECS and MECS instruments, as suggested by the following considerations. 
The misalignment of the X-ray telescopes with respect to the spacecraft Z axis was 
accurately calibrated using a set of dedicated observations (Matteuzzi \cite{alessio}) 
during the {\it Beppo}SAX Science Verification Phase. In the case of the MECS, the 
errors in the angles of the computed misalignment matrices $\mathrm{M_{mis}}$ (see the Appendix for details) 
are $\sim 15''$ (Chiappetti, priv. comm.).

Since this set of observations was carried out in 1996, the calibration of LECS and MECS misalignment has 
been based on attitude data affected by significant inaccuracies. Our conclusion is that this fact has induced a 
systematic error in the misalignment matrices computation of the two instruments.


\section{Correction of LECS and MECS coordinates}
\label{equations}

The observed dependence of LECS and MECS $\Delta \alpha$ and $\Delta \delta$ values on 
the spacecraft roll angle $\rho$ (see Fig.~\ref{roll}) allows us to parameterize, as a function 
of the roll angle, the LECS and MECS residual misalignment and to compute analytical formulae 
that can be used to correct the sky coordinates derived from the two instruments.

To this end, we have fitted the data plotted in Fig.~\ref{roll} with the function
$f(\rho) = A~\mathrm{cos}\,(\rho+\phi)+c$. We obtained the 
following best fits for LECS and MECS respectively:
\begin{eqnarray}
  \label{ra_lecs}
  \Delta \alpha_\mathrm{L}=15~\mathrm{cos}\,(\rho-140)+2 ~~~~~ \mathrm{[arcsec]} \\
  \Delta \delta_\mathrm{L}=19~\mathrm{cos}\,(\rho-236)-4 ~~~~~\, \mathrm{[arcsec]}
  \label{dec_lecs}
\end{eqnarray}
\begin{eqnarray}
  \label{ra_mecs}
  \Delta \alpha_\mathrm{M}=27~\mathrm{cos}\,(\rho-12)+3 ~~~~~~\, \mathrm{[arcsec]} \\
  \label{dec_mecs}
  \Delta \delta_\mathrm{M}=26~\mathrm{cos}\,(\rho-100)-4 ~~~~~\, \mathrm{[arcsec]}
\end{eqnarray}
where $\rho $ is the median of the roll angle values during the considered observation 
expressed in degrees.

These best fits are plotted as solid lines in Fig.~\ref{roll}. In Table~\ref{errors} we report the 
corresponding reduced $\chi^2$ values, the parameter best fit values and the associated errors 
at the 90\% confidence level for three interesting parameters.

From the definition of $\Delta \alpha$ and $\Delta \delta$ it follows that the corrected 
values of LECS and MECS coordinates are given by:
\begin{eqnarray}
  \label{ra_cor}
  \alpha^\mathrm{cor}_\mathrm{L,M} = \alpha_\mathrm{L,M} + \Delta \alpha_\mathrm{L,M} / 
                                 \mathrm{cos}\,\delta ~~\\
  \label{dec_cor}
  \delta^\mathrm{cor}_\mathrm{L,M} = \delta_\mathrm{L,M} + \Delta \delta_\mathrm{L,M} ~~~~~~~~~~~
\end{eqnarray}
Equations (\ref{ra_cor}) and (\ref{dec_cor}) can be used to correct the sky 
positions derived from LECS and MECS images. Alternatively these formulae can be 
applied to modify the headers' keywords of LECS and MECS FITS event files (see Sect.~\ref{event}).

In order to estimate the positioning accuracy of the two instruments after the 
coordinate correction, we have applied Eqs.~(\ref{ra_cor}) and (\ref{dec_cor}) 
to the coordinates of the 72 X-ray detections of our sample and computed again 
the offsets between LECS/MECS and optical positions.

The results are shown in Fig.~\ref{circles} (panels b and d). As can 
be seen, the offsets are significantly reduced and 
the non uniform distribution of $\Delta \alpha$ and $\Delta \delta$ values for the MECS 
is no longer present.

The 68\% and 90\% error radius values for LECS and MECS after the coordinate 
correction are reported in Table~\ref{tab} (bottom), where we can see that their positioning accuracy 
has improved by a factor of $\sim 2$. 

The residual error in the location of X-ray sources (16'' and 17'', for LECS and MECS respectively, at 
the 90\% confidence level) can be ascribed to the statistical uncertainty due to the Point Spread Function 
of the instruments (Fiore et al. \cite{fiore01}). Moreover, we outline that these values are of the same 
order (or even better) of the original LECS and MECS pixel size, showing that actually we have reached a 
positioning accuracy which is close to the limit of the detectors.


\begin{table} [h]
   \caption{Reduced $\chi^2$ and best fit parameters values (the associated errors are 
                 at the 90\% confidence level for three interesting parameters) relative to the best 
                 fits of Eqs.~(\ref{ra_lecs}), (\ref{dec_lecs}), 
                 (\ref{ra_mecs}) and (\ref{dec_mecs}).
   }
   \label{errors}
     \begin{tabular}{lcccc}
     \hline
     \hline
&	&	 &                   &                   \\
Data Set~~ & $\chi^2_{\nu}$\,(d.o.f.) & ~~~$A$ ['']~~ & ~~~$\phi$ [$^\circ$]~~ & ~~~$c$ ['']~~ \\
           &	                       &   	      &            &          \\
     \hline
&	&	 &                   &                   \\
$\Delta \alpha_\mathrm{L}$ & 1.43\,(69) & $15\pm3$ & $-140\pm11$ & $~~2\pm2$ \\
$\Delta \delta_\mathrm{L}$ & 1.06\,(69) & $19\pm3$ & $-236\pm12$ & $-4\pm2$ \\
$\Delta \alpha_\mathrm{M}$ & 1.05\,(69) & $27\pm3$ & $ -12\pm 5$ & $~~3\pm2$ \\
$\Delta \delta_\mathrm{M}$ & 1.34\,(69) & $26\pm2$ & $-100\pm 6~$ & $-4\pm2$ \\
&	&	 &                   &                   \\
     \hline
     \end{tabular}
\end{table}



\subsection{Off-axis sources}
\label{off-axis}

We have seen that LECS and MECS images are shifted in RA and Dec by an amount of $\sim$ 20''.

Our analysis is based on the positions of on-axis sources and therefore it is not capable to check if a 
further rotation of the X-ray images around the center of the field of view is present. However, as the 
following argument shows, the effect on the coordinates of off-axis sources is expected to be negligible.

The amplitude of the rotation should be $\sim$ 30'', i.e. the amount of 
the spacecraft attitude inaccuracy. A 30'' rotation of the image around its center has 
the effect of shifting the coordinates of a source located at an off-axis angle of 30 arcmin 
of a very small distance ($\sim$ 0.3'') and can be therefore neglected.

We have verified this comparing the {\it Beppo}SAX X-ray and 
optical positions of a small sample of off-axis sources. In all cases we have 
found that Eqs.~(\ref{ra_cor}) and (\ref{dec_cor}) corrects properly the 
positions of the off-axis sources.


\section{Positioning accuracy before May 1998}
\label{May98}

In this section we discuss the positioning accuracy of LECS and MECS for observations 
carried out before May 6th 1998.

We restricted first our study to a sample of 77 X-ray fields relative to the time interval 
June 1996-May 1997.
The sample consists of 50 pointings of AGNs, 21 of X-ray binaries and 6 of Stars, and 
has been selected according to the first three criteria listed in Sect.~\ref{data}. For the 
LECS we considered a subsample of 57 images, since in 20 of the selected fields the exposures were 
very short or there was a lack of data. Since the MECS1 unit was still working during this period we 
used co-added MECS1+MECS2+MECS3 images. 

We repeated for LECS and MECS fields the analysis described in the previous sections. We found a larger 
scatter of the offsets between the X-ray and optical positions with respect to observations performed after 
May 6th 1998, as expected from the fact that the spacecraft attitude data for this sample are less 
accurate. 


\begin{table} [h]
   \caption{LECS and MECS (three units) 68\% and 90\% error radius values before and after the coordinate 
correction of Eqs.~(\ref{ra_cor}) and (\ref{dec_cor}) for observations carried out in the period 
June 1996-May 1997.}
   \label{tab2}
     \begin{tabular}{cccc}
     \hline
     \hline
  &   ~Instrument~ & ~68\% radius~ & ~90\% radius~ \\
     \hline
	&	 &                   &                   \\
before   &  LECS & 30'' & 41'' \\
~correction~   &  MECS & 37'' & 53'' \\
	&	 &                   &                   \\
     \hline
	&	 &                   &                   \\
after   &  LECS & 21'' & 29'' \\
correction   &  MECS & 19'' & 38'' \\
	&	 &                   &                   \\
     \hline
     \end{tabular}
\end{table}


In Table \ref{tab2} (top) we report the corresponding LECS and MECS 68\% and 90\% error radius 
values (numbers have been rounded to unity).

We next verified that Eqs.~(\ref{ra_lecs}), (\ref{dec_lecs}), (\ref{ra_mecs}) and 
(\ref{dec_mecs}) describe well the dependence of $\Delta \alpha$ and $\Delta \delta$ on the 
roll angle and can be therefore used to correct the LECS and MECS celestial coordinates also for 
observations carried out during this time interval (the corresponding reduced $\chi^2$ values are 
$\chi^2_{\nu} = 1.69$ for $\Delta \alpha_{\mathrm L}$, $\chi^2_{\nu} = 1.78$ for $\Delta \delta_{\mathrm L}$, 
$\chi^2_{\nu} = 1.42$ for $\Delta \alpha_{\mathrm M}$ and $\chi^2_{\nu} = 1.58$ for $\Delta \delta_{\mathrm M}$).

The separate analysis of MECS1, MECS2 and MECS3 images confirmed that, also for the 
observations of this sample, there is no difference in the misalignment of the single units.

In Table \ref{tab2} (bottom) we list the LECS and MECS 68\% and 90\% error radius values after the coordinate 
correction of Eqs.~(\ref{ra_cor}) and (\ref{dec_cor}) for the 77 (57 for the LECS) X-ray detections of the sample.

As can be seen, the correction improves the source location accuracy of both instruments. Moreover, from 
a comparison of the error radii of Table \ref{tab2} with those of Table \ref{tab}, we see that the inaccurate 
attitude data during this period affect significantly the positioning precision of LECS and 
MECS.

We have not repeated the whole analysis for observations performed between June 1997 and May 6th 1998, since 
the rather large spacecraft attitude inaccuracies (up to 1-2 arcmin) during this time interval dominate 
the $\sim$ 20'' residual misalignment of LECS and MECS.

Given the attitude inaccuracy, a 90\% error radius of $\sim$ 1.2 arcmin should be used for LECS and MECS 
observations performed during this period. A boresight correction with respect to the known position of a source 
in the field of view (if available) may be used to greatly reduce this error.


\section{Correction of LECS and MECS event files}
\label{event}

Equations~(\ref{ra_cor}) and (\ref{dec_cor}) correct the sky positions derived from{\it Beppo}SAX  
LECS and MECS data. In particular the accuracy of all the celestial coordinates already published in literature 
can be improved applying the two formulae.

These equations can also be used to modify the headers of LECS and MECS FITS event files, thus allowing to derive 
from data sky positions that have the coordinate correction already applied.

To this end, we have developed a specific task that updates according to Eqs.~(\ref{ra_cor}) and (\ref{dec_cor}) 
the values of the sky coordinates at reference pixel stored in the headers of LECS and MECS event files.

The task, named {\it saxposcor}, can be downloaded from a dedicated web page where a detailed description of the task 
and of its usage can be found (http://www.asdc.asi.it/bepposax/coord\_correction.html).

The headers of all LECS and MECS event files in the {\it Beppo}SAX archive at the ASDC have been modified 
with {\it saxposcor} on March 12 2002 and are available on-line. The {\it Beppo}SAX data reduction 
software SAXDAS (since version 2.3.0) includes the coordinate correction discussed here and 
therefore automatically corrects the headers of LECS and MECS event files.

All the details concerning the the {\it Beppo}SAX ASDC archive correction can be found on the mentioned web page.


\section{Summary}
\label{summary}

We have presented a study of the positioning accuracy of the {\it Beppo}SAX LECS and MECS instruments using 
a set of archival images for which the position of the target is known to within 1''.

We have found that a residual misalignment with respect to the spacecraft Z axis of $\sim 17''$ for the LECS 
and $\sim 27''$ for the MECS was present.

This residual misalignment introduces a systematic error in the derived LECS and MECS source positions 
which depends on the spacecraft roll angle. 

We give analytical formulae that can be used to correct the celestial coordinates derived from the two instruments. 
After the coordinate correction the source location error of LECS and MECS reduces respectively to 16'' and 17'' 
at the 90\% confidence level. This represents an improvement by a factor of $\sim 2$. The residual uncertainty 
can be ascribed to the instrumental Point Spread Function and is of the order of the detectors pixel size.

Our work constitutes the first joint presentation of LECS and MECS imaging capabilities. 
The improvement presented here, which can be considered as a final reference of the location 
accuracy of LECS and MECS, can be useful for multi-wavelength studies and future works based on 
{\it Beppo}SAX source catalogs.

\begin{acknowledgements}

We thank F. Fiore and P. Giommi for useful discussions and a careful supervision of the work, Tim Oosterbroek 
for having stimulated the beginning of this study and the referee, L. Chiappetti, for helpful comments 
which improved the paper. We also thank B. Saija and F. Tamburelli for their help in the development of 
the script {\it saxposcor}.

\end{acknowledgements}


\appendix

\section{On Satellite Attitude}
\label{app}

In this Appendix we give some definitions and formulae concerning the {\it Beppo}SAX attitude data, the 
instruments misalignment and the conversion of detector coordinates to celestial coordinates.

\subsection{Spacecraft Attitude}

The satellite attitude is defined as the orientation of the satellite axes (X,Y,Z) as a function of time. 

The attitude is controlled through three on-board optical star-trackers which, for a given pointing, 
image the available guide stars in their field of view. When the guide stars 
are unavailable (e.g. for Earth occultation) the satellite orientation is determined by a set of additional 
instruments, i.e. by on-board gyroscopes, a solar sensor and a magnetic field sensor.

The guide stars celestial coordinates and the misalignment matrices of the three star-trackers with respect 
to the spacecraft axes are subsequently used on ground, by the {\it Beppo}SAX Attitude and Orbit 
Control Subsystem (AOCS), to reconstruct the spacecraft attitude. 

Specifically, the attitude data consist of a file containing RA and Dec values of the three spacecraft axes 
($\alpha_\mathrm{X,Y,Z}$ and $\delta_\mathrm{X,Y,Z}$) as a function of time (the time bin is 0.5 s). 

The satellite attitude may be also described in terms of a rotation with respect to the celestial coordinate 
system. Following the Euler angles definitions (Goldstein \cite{Goldstein}) we obtain the following 
expressions for the three rotation angles $\Phi$, $\Theta$ e $\Psi$:
\begin{eqnarray}
  \label{euler1}
  \Phi = 90 + \alpha_\mathrm{Z}   ~~~~~~~~~~~~~~~ \\
  \label{euler2}
  \Theta = 90 - \delta_\mathrm{Z} ~~~~~~~~~~~~~~~\, \\
  \label{euler3}
  \Psi = \mathrm{atan}\,(\mathrm{sin}\,\delta_\mathrm{X}/\mathrm{sin}\,\delta_\mathrm{Y}) 
\end{eqnarray}

Given the attitude Euler angles it is useful to define the roll angle $\rho$:
\begin{equation}
  \rho = \Psi -90
  \label{roll_def}
\end{equation}
and to describe the spacecraft attitude with the three angles  $\alpha_\mathrm{Z}$, $\delta_\mathrm{Z}$ and 
$\rho$.

\subsection{Instruments Misalignments}

Given a satellite attitude reconstruction, in order to associate a celestial coordinates grid to the LECS 
and MECS images the misalignment between the optical axes of the X-ray telescopes 
with respect to the satellite axes (X,Y,Z) has to be taken into account.

This misalignment is described in terms of three subsequent rotations around the satellite axes.
If we call $\mathrm{M_{mis}}$ the instruments misalignment matrix and $\mathrm{M_{att}}$ the spacecraft 
Euler matrix associated to the angles $\Phi$, $\Theta$ e $\Psi$, we can define the matrix $\mathrm{M_{det}}$
\begin{equation}
 \mathrm{M_{det}} = \mathrm{M_{mis}} ~ \mathrm{M_{att}}
  \label{detector}
\end{equation}
which allows to associate a celestial coordinate system to the detector image. The misalignment 
matrices $\mathrm{M_{mis}}$ have been computed by the LECS and MECS scientific teams.

\subsection{Sky Coordinates Computation}

A detailed discussion of the procedure used to convert LECS and MECS raw detector pixels to linearized 
detector pixels and then to sky pixels is beyond the scope of this Appendix and we briefly summarise 
only the main steps.

First, the raw (electronic) detector pixels are converted to a physical distance (in mm) on the detector 
surface. Since the response of the detector is affected by some nonlinearities, which have been calibrated on 
ground, this step involves a coordinate linearization which corrects image distortions.

Second, the distance in mm on the detector is converted to detector linearized pixels. A choice of a pixel 
size of 8'', for both LECS and MECS, is included in this step.

Third, detector linearized pixels are converted to sky coordinates pixels. This step involves the use of 
the matrix $\mathrm{M_{det}}$ which allows to associate celestial coordinates to the detector linearized image. 

The celestial coordinates of a ``reference'' detector pixel and  the roll angle, computed as the median of 
the values assumed during the observation, are stored in the headers of LECS and MECS event files.



\begin{thebibliography}{}

   \bibitem[2000]{lucio} Antonelli, L.A., Piro, L., Vietri, M., et al. 2000, ApJ, 545, 39

   \bibitem[1997a]{boella97a} Boella, G., Butler, R.C., Perola, G.C., et al. 1997a, A\&AS, 122, 299

   \bibitem[1997b]{boella97b} Boella, G., Chiappetti, L., Conti, G., et al. 1997b, A\&AS, 122, 327

   \bibitem[2001]{comastri} Comastri, A., Fiore, F., Vignali, C., et al. 2001, MNRAS,327, 781

   \bibitem[2001]{feroci} Feroci, M., Antonelli, L.A., Soffitta, P., et al. 2001, A\&A, 378, 441

   \bibitem[1999]{fiore99} Fiore, F., La Franca, F., Giommi, P., et al. 1999, MNRAS,306, 55

   \bibitem[2001]{fiore01} Fiore, F., Giommi, P., Vignali, C., et al. 2001, MNRAS, 327, 771

   \bibitem[1992]{giommi92} Giommi, P., Angelini, L., Jacobs, P., \& Tagliaferri, G. 1992, A.S.P. 
                            Conference Series, Vol. 25, p. 100

   \bibitem[1997]{pao_fab} Giommi, P., Fiore, F. 1997, Data Analysis in Astronomy IV, p.93

   \bibitem[1998]{giommi98} Giommi, P., Fiore, F., Ricci, D., et al. 1998, Nucl. Phys. B, 69/1-3, 691

   \bibitem[2000]{giommi00} Giommi, P., Perri, M., \& Fiore, F. 2000, A\&A, 362, 799

   \bibitem[1980]{Goldstein} Goldstein, H., 1980, {\it Classical Mechanics}, Addison-Wesley

   \bibitem[2001]{zand01} in 't Zand, J.J.M., Kuiper, L., Amati, L., et al. 2001, ApJ, 559, 710

   \bibitem[2002]{zand02} in 't Zand, J.J.M., Miller, J.M., Oosterbroek, T., \& Parmar, A.N. 2002, A\&A, 
                        submitted (astro-ph/0205535)

   \bibitem[2002]{lafranca} La Franca, F., Fiore, F., Vignali, C., et al. 2002, ApJ, 570, 100

   \bibitem[1998]{alessio} Matteuzzi, A. 1998, PhD Thesis, Padova University

   \bibitem[1997]{parmar} Parmar, A.N., Martin, D.D.E., Bavdaz, M., et al. 1997, A\&AS, 122, 309

   \bibitem[2002]{piro}  Piro, L., Frail, D.A., Gorosabel, J., et al. 2002, ApJ, in press, 
                         (astro-ph/0201282)

   \bibitem[1998]{ricci} Ricci, D., Fiore, F., \& Giommi, P. 1998, Nucl. Phys. B, 69/1-3, 618

   \bibitem[1992]{turon} Turon, C., Creze, M., Egret, D., et al. 1992, ESA SP-1136

   \bibitem[1995]{xrbcat} van Paradijs, J., "X-ray Binaries", eds. W.H.G. Lewin, 
                          J. van Paradijs, and E.P.J. van den Heuvel, 
                          Cambridge University Press, 1995.

   \bibitem[1996]{veron} Veron-Cetty, M.-P. \& Veron, P. 1996, A\&AS, 115, 97

   \bibitem[2001]{vignali} Vignali, C., Comastri, A., Fiore, F., La Franca, F. 2001, A\&A, 370, 900

\end{thebibliography}
\end{document}